# Dual-Mode SERS and Colorimetric Sensor for Lung Cancer VOC-Biomarker Detection Using Hydrogel Patches

Jialin Li [b, a], Shaowei Liu [b], Amil Aligayev [a, c, e], Hao Jiang [a, d], Muhammad [a], Xin Yu [a, e], Jin Tao [f], Jiaqi Wang [a, d], Agnieszka Jastrzębska [g], Qing Huang [a, e*]

[a] CAS Key Laboratory of High Magnetic Field and Iron Beam Physical Biology, Institute of Intelligent Machines, Hefei Institute of Physical Sciences, Chinese Academy of Sciences, Hefei, 230031, China

[b] College of Integrated Circuits, Southeast University, Nanjing, 214026, China

[c] NOMATEN Centre of Excellence, National Centre for Nuclear Research, 05-400 Otwock, Poland

[d] School of Life Sciences, Anhui Agricultural University, 130 Changjiang West Road, Hefei, 230036, China

[e] Science Island Branch of Graduate School, University of Science and Technology of China, Hefei, 230026, China

[f] School of Metallurgy Engineering, Anhui University of Technology, Ma'anshan 243032, Anhui, China

[g] Warsaw University of Technology, Faculty of Mechatronics, św. Andrzeja Boboli 8, Warsaw 02-525, Poland

*: **Corresponding author:** huangq@ipp.ac.cn (Q. Huang)

## Abstract

Hexanal, a volatile organic compound (VOC), is a potential biomarker for the early detection of lung cancer. In this study, we developed a dual-mode flexible biosensor that integrates surface-enhanced Raman scattering (SERS) and colorimetric detection for the quantitative analysis of hexanal in human exhaled breath. The biosensor employs Ag nanocubes wrapped with Co-Ni layered double hydroxide (AgNCs@Co-Ni LDH) as a functional matrix, offering both superior SERS enhancement and efficient VOC adsorption properties. To achieve selective detection, AgNCs@Co-Ni LDH were incorporated into agarose hydrogels along with 3-methyl-2-benzothiazolinone hydrazone (MBTH). The resulting MBTH-AgNCs@Co-Ni LDH/hydrogel-patch facilitate the oxidation of hexanal, enabling simultaneous colorimetric and SERS signal generation, while producing acrizine, a blue-colored reaction product. This hydrogel-based dual-mode sensing platform exhibits high selectivity, excellent stability, and precision in SERS-based hexanal detection. The detection limit for the SERS method was determined to be as low as $3.34\times10^{-13}$ M. Furthermore, developed and optimized compact CNN-based multi-terminal intelligent recognition system for enhanced hydrogel-patch detection through AI-driven, portable, and real-time colorimetric analysis. Therefore, this work not only enables the effective detection of hexanal in the exhaled breath of suspected lung cancer patients, underscoring its potential for early lung cancer screening, but also establishes a foundation for the development of multimodal hydrogel biosensors for broader applications in disease diagnosis.

## 1. Introduction

The early and accurate detection of lung cancer remains a critical challenge in clinical oncology. Owing to its non-specific symptoms in the initial stages, lung cancer is frequently diagnosed at an advanced stage, where therapeutic options are limited, and prognosis is poor. Current diagnostic methods such as chest X-rays, sputum cytology, low-dose spiral computed tomography (LDCT), and positron emission tomography (PET) have contributed to improved screening [1,2]. However, these techniques are often costly, time-intensive, and require specialized infrastructure and personnel [3]. Therefore, the development of non-invasive, rapid, sensitive, and cost-effective diagnostic tools is of paramount importance for facilitating early-stage lung cancer screening and improving patient outcomes [4,5].

In recent years, breathomics has emerged as a promising non-invasive strategy for disease detection, particularly through the analysis of volatile organic compounds (VOCs) present in exhaled breath [6]. VOCs have demonstrated great potential as biomarkers for lung cancer due to their metabolic origin and differential expression between healthy individuals and patients [7–9]. Gas chromatography (GC) and gas chromatography-mass spectrometry (GC-MS) are currently the most reliable techniques for VOC analysis [10,11]. Nevertheless, these methods are hindered by complex sample preparation, high operational costs, and dependence on advanced equipment.

Hexanal, a type of aldehyde, has attracted significant attention among potential biomarkers due to its notable specificity for lung cancer [12]. Unlike other straight-chain aldehydes commonly found in a range of pulmonary disorders, hexanal exhibits minimal cross-reactivity and is significantly elevated in the breath of lung cancer patients compared to healthy individuals [13–15]. While hexanal serves as a promising diagnostic marker, its accurate detection at low concentrations in exhaled breath poses significant analytical challenges [14,16].

To address these challenges, recent efforts have focused on optical sensing technologies, which offer high sensitivity, rapid response times, and operational simplicity attributes well-suited for point-of-care (POC) diagnostics. However, traditional optical sensors often rely on single-signal output, making them susceptible to environmental and instrumental interferences [17,18]. In contrast, multimodal sensing strategies that integrate different transduction mechanisms offer enhanced accuracy, reliability, and adaptability to real-world conditions [19].

In this context, a dual-mode biosensor that combines surface-enhanced Raman scattering (SERS) and colorimetric sensing could be a profound solution. For example, SERS offers ultra-high sensitivity of analytes through enhancement of Raman signals with target-specific recognition, while colorimetry provides visually intuitive outputs and simple instrumentation [20–23]. The synergistic combination of these modalities enables cross-validation of signals generated through independent physical mechanisms, significantly reducing false positives and improving diagnostic confidence [24,25]. However, most dual-mode sensors rely on inflexible substrates or colloidal substrates, which limit their portability, reusability, and ease of integration into wearable formats. To overcome these limitations, a 3D-hydrogel matrix could be an excellent alternative to host the SERS and colorimetric nano reagents. Hydrogels are ideal candidates for wearable biosensing platforms due to their flexibility, high porosity, biocompatibility, and ability to maintain a hydrated microenvironment [26,27].

In this study, we developed a dual-mode detection strategy based on 3D-hydrogel stabilized Ag nanocubes supported on Co-Ni layered double hydroxide (AgNCs@Co-Ni LDH), preventing nanoparticle aggregation and preserving their high SERS activity over time. The AgNCs@Co-Ni LDH nanocomposites were chosen for their excellent plasmonic properties, chemical stability, and synergistic catalytic behavior, which collectively enhanced both Raman signal amplification and colorimetric reactivity. Functionalization of the hydrogel with an optimal concentration of 3-methyl-2-benzothiazolinone hydrazone (MBTH) aqueous solution allowed it to act as a reaction reservoir, creating an ideal microenvironment for the rapid reaction between MBTH and hexanal, leading to notable colorimetric changes and distinguishable SERS signals. The synthesis procedure of the hydrogel-based biosensor is illustrated in **Scheme 1A**. The process started from the formation of AgNCs, which were subsequently encapsulated by ZIF-67 to form AgNCs@ZIF-67. This composite was then subjected to an etching process using $Ni^{2+}$ ions, resulting in the formation of AgNCs@Co-Ni LDH. The nanocomposite was then incorporated into an agarose hydrogel matrix along with MBTH to fabricate the hydrogel-patch, a dual-mode sensing platform for colorimetric and SERS detection. **Scheme 1B** illustrates the detection mechanism, where volatile hexanal could interact with MBTH within the hydrogel matrix to produce a blue-colored derivative, while also generating SERS signals via the AgNCs@Co-Ni LDH. This integration of selective chemistry within a stable hydrogel framework ensured quick, reproducible, and sensitive detection of hexanal. To enhance usability, we also incorporated smartphone-based image acquisition and low-cost 3D-printed accessories, enabling real-time analysis and on-site monitoring. To support automated and accurate analysis, a multi-terminal detection algorithm based on deep learning was implemented. Among various models tested, the convolutional neural network (CNN) architecture outperformed traditional machine learning approaches in classifying subtle color changes. By training a task-specific CNN, we enabled fast and reliable recognition of hydrogel color transitions on both personal computers and smartphones, offering a practical and accessible platform for the early detection of lung cancer biomarkers. **Scheme 1C** presents the conceptual clinical application, showing the hydrogel-patch integrated into a wearable device for the early, non-invasive detection of lung cancer through breath analysis, and **Scheme 1D** shows the full diagnostic workflow of the biosensor system, which starts with breath sample collection, followed by colorimetric reaction, image acquisition, deep learning-based analysis, and ultimately enables accurate and rapid lung cancer screening. As such, by harnessing the advantages of AgNCs@Co-Ni LDH supported by SERS and colorimetric platform, our biosensing approach sought to provide new solutions to the field of precision biomedicine and personalized diagnostics.

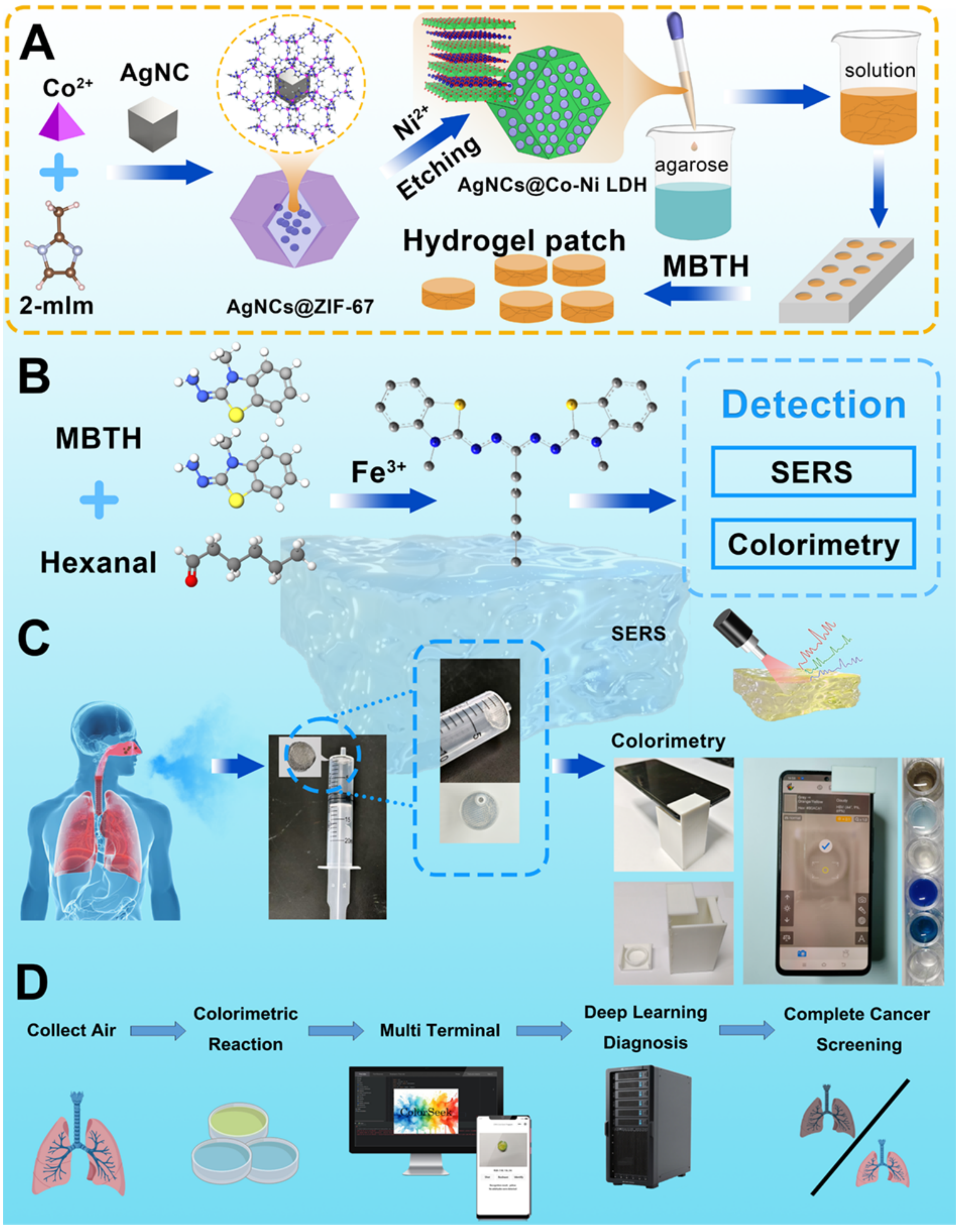
A
Co2+
AgNC
2-mIm
AgNCs@ZIF-67
Ni2+
Etching
AgNCs@Co-Ni LDH
agarose
solution
Hydrogel patch
MBTH
B
MBTH
Hexanal
Fe3+
Detection
SERS
Colorimetry
C
SERS
Colorimetry
D
Collect Air
Colorimetric
Reaction
Multi Terminal
Deep Learning
Diagnosis
Complete Cancer
Screening

**Scheme 1.** (A) Synthesis pathway of AgNCs@LDH hydrogel-patches, illustrating the sequential formation of AgNCs, AgNCs@ZIF-67, and AgNCs@LDH. (B) Schematic representation of the detection mechanism for hexanal as VOCs using the dual-mode hydrogel-based biosensor. (C) Conceptual illustration of the hydrogel-based biosensor's application in the early clinical screening of lung cancer patients. (D) Schematic diagram of lung cancer screening based on deep learning, from respiratory gas collection and colorimetric reaction to multi-terminal diagnosis.

## 2. Experimental section

### 2.1 Reagent

All chemicals were of analytical grade and used without further purification. Silver nitrate ($AgNO_3$, ≥ 99.8%), poly(vinylpyrrolidone) (PVP, K30), ethylene glycol, sodium sulfide ($Na_2S$, 98.0%), methanol (≥ 99.5%), ethanol (≥ 99.7%), propionaldehyde (≥ 99.5%), cobalt(II) nitrate hexahydrate (Co(NO$_3$)2·6H$_2$O, ≥ 99.0%), nickel(II) nitrate hexahydrate ($Ni(NO_3)_2 \cdot 6H_2O$, ≥ 98.0%), 2-methylimidazole (99%), benzaldehyde (≥ 97.0%), salicylaldehyde (98%), and 3-methyl-2-benzothiazolinone hydrazone hydrochloride monohydrate were purchased from Sinopharm, Aladdin, and Sangon Biotech Co., Ltd (China). Agarose hydrogel powder was obtained from Biofroxx (Germany). Ultrapure water (18.25 MΩ·cm) was used in all procedures. All glassware and magnetic stirrers were cleaned with aqua regia and thoroughly rinsed with ultrapure water before use.

### 2.2 Preparation of AgNCs@Co-Ni LDH SERS substrates

AgNCs were synthesized via sulfide-mediated polyol method and subsequent redispersion in methanol. In brief, a ZIF-67 precursor solution containing [$Co(NO_3)_2 \cdot 6H_2O$] and 2-methylimidazole was prepared, into which the AgNCs were immediately introduced to initiate the growth of AgNCs@ZIF-67. The resulting AgNCs@ZIF-67 composite served as a sacrificial template for the fabrication of AgNCs@Co-Ni LDH through an etching process in $Ni(NO_3)_2 \cdot 6H_2O$ solution. The synthesized AgNCs@Co-Ni LDH was stored in the dark at 4 °C to preserve its structural integrity. Additional experimental details are available in the Supporting Information.

### 2.3 Preparation of MBTH-AgNCs@Co-Ni LDH/hydrogel-patches

To prepare the hydrogel matrix, 0.5 g of agarose was added to 50 mL of deionized water and heated to 100 °C until fully dissolved. Subsequently, 50 mg of MBTH and 2 mL of AgNCs@Co-Ni LDH nanoparticles were introduced into the solution. The mixture was then stirred at 900 rpm at 90 °C for 20 min to ensure homogeneous functionalization of the hydrogel network. Afterwards, the solution was allowed to cool and cast into a pre-designed mold fabricated via 3D printing (**Figure S1**). The resulting hydrogel-patches were sealed with a protective film and stored in the dark at 4 °C for future use.

## 2.4 Sample preparation and dual-mode response study

A single hydrogel-patch h was placed in a sealed container alongside liquid aldehyde samples of varying concentrations, ensuring no direct contact between the patch and the liquid phase. The system was incubated at 80 °C for one hour to promote aldehyde volatilization and diffusion. Following incubation, 5 μL of a 1% ferric chloride was applied to the hydrogel-patch to induce the formation of aldehyde-MBTH derivatives. Following a reaction time of 5 min, Raman spectra were acquired from the patch to evaluate the system's sensing performance. For each sample, spectra were collected in triplicate to ensure reproducibility. Additionally, colorimetric responses were quantified using a smartphone-based color analysis application (Color Grab). Detailed experimental procedures and characterization conditions are provided in the Supporting Information.

## 2.5 Real sample preparation and analysis

The performance and reliability of the developed dual-mode biosensor were further validated through analysis of exhaled breath samples collected from confirmed lung cancer patients and healthy controls. All procedures were conducted in accordance with ethical guidelines and approved by the Ethics Committee of the First Affiliated Hospital of Anhui Medical University (Approval No. 84230021). Informed consent was obtained from all participants prior to enrollment. Participants were required to fast for at least 8 h prior to sample collection to minimize dietary influences on VOCs. Sampling was performed by having subjects wear a nasal clip and exhale deeply into pre-conditioned Tedlar® gas sampling bags (5 L capacity) via a disposable mouth mask. Mixed expiratory breath samples were collected without restricting to specific respiratory phases to capture representative VOC profiles. In select cases, subjects were monitored continuously using face masks embedded with hydrogel-patches, enabling direct in situ detection. All breath samples were collected by trained personnel employing pre-cleaned Tedlar® bags that had undergone baking at 60 °C for 3 h the night before use to remove residual contaminants, followed by multiple purges with high-purity nitrogen gas. To maintain environmental control and ensure data validity, breath collection occurred in contamination-free settings, with simultaneous ambient air samples collected as controls. Prior to sampling, participants rinsed their mouths with purified water, performed deep nasal inhalation, and exhaled fully into the Tedlar® bags, accumulating a total breath volume of approximately 5 L. The collection and handling procedures adhered strictly to established respiratory sampling protocols [28,29], ensuring consistency and reproducibility across the samples.

## 2.6 Characterization of AgNCs@Co-Ni LDH

Structural, morphological, and chemical properties of the sensing materials were systematically analyzed using a suite of advanced characterization techniques. The crystalline phases were identified via powder X-ray diffraction (X'Pert PRO, PANalytical, Netherlands) using Cu Kα radiation (λ = 1.5416 Å), with diffraction patterns recorded over a 2θ range of 5° to 85° at a scanning step of 0.02°. Surface morphology and microstructure were examined by field-emission scanning electron microscopy (FE-SEM, S-4800, Hitachi, Japan) and high-resolution transmission electron microscopy (HR-TEM, JEM-2100F, JEOL, Japan) operating at 200 kV. Elemental composition and spatial distribution were further analyzed using energy-dispersive X-ray spectroscopy (EDX) integrated with SEM (Sirion 200, Fei, USA). Surface chemical states were investigated by X-ray photoelectron spectroscopy (XPS, ESCALAB 250Xi, Thermo Fisher Scientific, USA) using an Al Kα source (hν = 1486.6 eV). Vibrational properties and molecular bonding features were assessed using Fourier transform infrared spectroscopy (FTIR, Alpha, Bruker, Germany). A portable Raman spectrometer (785 nm, 20 mW; Ruhai Optoelectronics) was used for SERS measurements, with an exposure time of 10 seconds. Extinction spectra were measured using a UV-vis spectrophotometer (UV-2550, Shimadzu, Japan) in the wavelength range of 300-800 nm.

## 2.7 Deep learning-based application for color recognition

In this study, we developed ColorSeek, a deep learning-based application designed to automatically classify colorimetric responses from hydrogel-patches. Advances in deep learning and computer vision provide promising solutions for medical image analysis [30]. By integrating computer vision techniques and CNN, the system facilitates rapid and accurate interpretation of sensor color changes associated with aldehyde detection. The application was implemented in Python, incorporating libraries such as *Tkinter* for graphical user interface (GUI) development, OpenCV for real-time image capture and processing, and *TensorFlow/Keras* for neural network deployment. The training dataset comprised labeled images of hydrogel-patches exposed to exhaled breath, categorized according to detection results. Specifically, we used bilinear interpolation to adjust the input image to a fixed 224×224 pixel size to match the input requirements of the model. Subsequently, the image was converted from PIL format to a NumPy array of type float32 shaped like (height, width, 3). Next, we divided the dataset into training set and test set according to an 8:2 ratio, and normalized the pixel values using Z-score normalization. During the training phase, we also applied data augmentation techniques, including random rotation and translation, to improve generalization capabilities. Given the proven effectiveness of CNNs in medical image classification tasks [31], we adopted a compact CNN architecture optimized using the Adam optimizer and the binary cross-entropy loss function. The model was trained over 30 epochs and achieved a final accuracy of over 98%. Following training, relevant metrics and visual outputs were saved, and the model was exported in .h5 format for later inference. During operation, the application loads the pre-trained model and initiates a GUI via Tkinter. Real-time video feed is handled by OpenCV, refreshing at 10-ms intervals. The interface includes three interactive options: image acquisition ("Take Photo"), reset ("Retake"), and prediction initiation ("Start Recognition"). Once an image is captured, it undergoes preprocessing before being fed into the CNN for classification. The prediction is based on a probability threshold of 0.5, and the corresponding color category is displayed. In cases of error or anomalies, the system triggers an appropriate alert. Upon exiting, all hardware resources and GUI components are appropriately released.

## 3. Results and discussion

### 3.1. Synthesis and characterization

The morphological and compositional characteristics of the synthesized materials were investigated using SEM, TEM, and elemental mapping techniques. As shown in **Figure 1A**, the SEM image of the hydrogel-patch reveals a highly porous three-dimensional network, which was favorable for the diffusion of target gas molecules and facilitates efficient interaction with embedded nanomaterials. Elemental mapping results (**Figure 1B**) further confirmed the uniform distribution of key elements, including C, O, Co, Ni, and Ag, across the hydrogel matrix. This uniform dispersion indicated the successful integration of AgNCs@Co-Ni LDH nanocomposites into the agarose-based hydrogel. Comparative SEM and elemental mapping images of AgNCs, AgNCs@ZIF-67 and AgNCs@Co-Ni LDH are provided in **Figure S2-S5**, further supporting the effective dispersion of the active components. TEM analysis of AgNCs@Co-Ni LDH (**Figure 1C**) revealed well-dispersed silver nanocubes embedded within the LDH structure, as indicated by the red-circled regions. These findings validate the structural integrity and nanoscale uniformity of the functionalized hydrogel system, essential for its application in sensitive and selective colorimetric sensing. Additional TEM micrographs of AgNCs and AgNCs@Co-Ni LDH are presented in Figure S6, with the images of AgNCs showing an average side length of 45 nm based on measurements of 50 particles.

X-ray diffraction analysis was conducted to elucidate the crystalline structure of agarose hydrogel-based AgNCs@Co-Ni LDH composites. As shown in **Figure 1D**, the AgNCs@Co-Ni LDH display prominent diffraction peaks at 11.6°, 23.5°, and 34.0°, corresponding to the (003), (006), and (009) crystallographic planes, respectively which are reflections are characteristic of the LDH structure. The XRD pattern of the hydrogel-patch retain all the characteristic diffraction peaks of LDH, even after the incorporation of agarose hydrogel and MBTH, which indicates that the crystallinity and structural integrity of the LDH framework are preserved during the functionalization process. In contrast, the XRD pattern of the agarose hydrogel exhibits a broad, diffused profile devoid of sharp crystalline peaks, which is an indicative of its inherently amorphous polymeric nature.

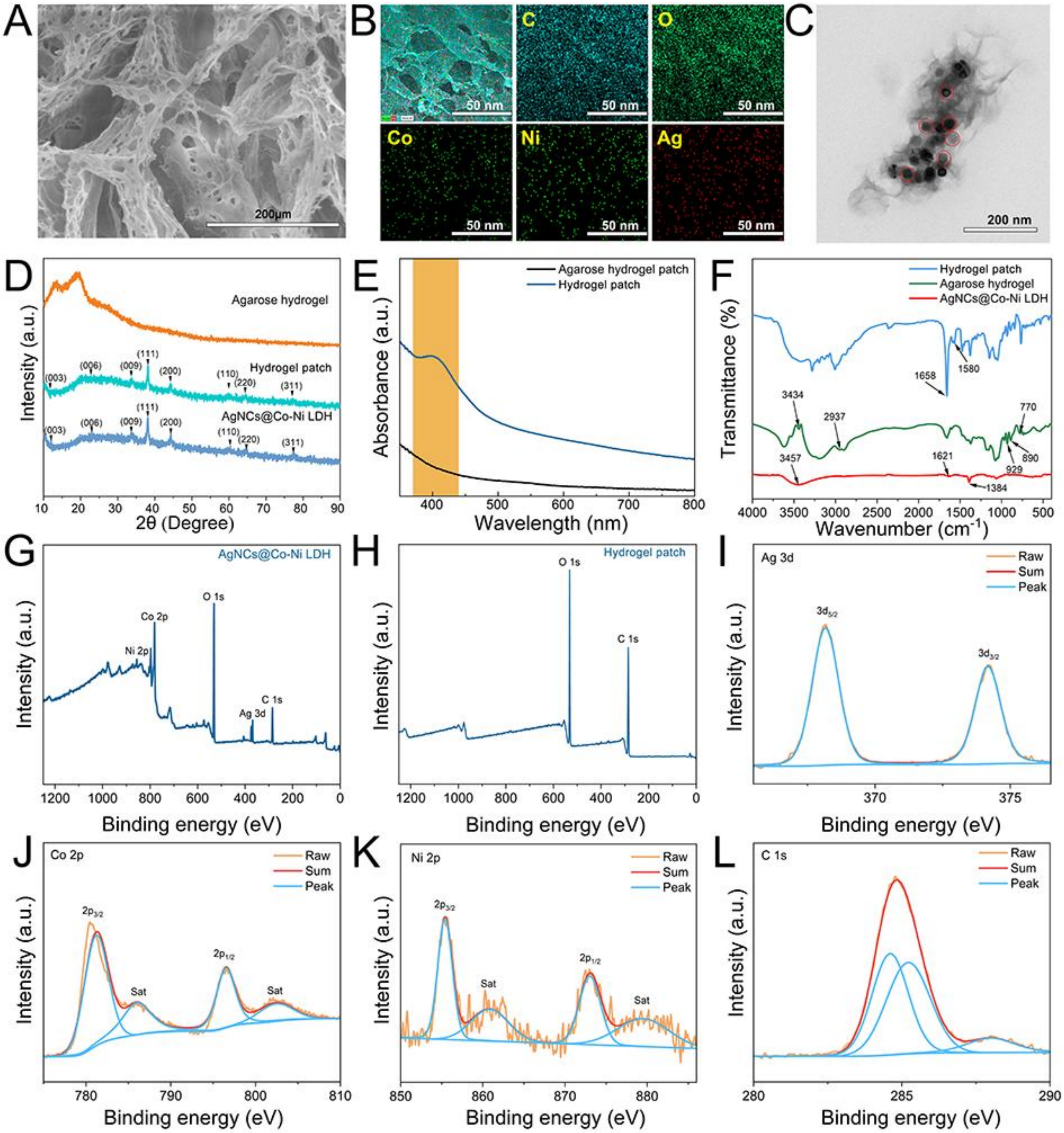

Figure 1. (A) SEM image of the hydrogel-patch. (B) SEM image of the hydrogel-patch with corresponding elemental mapping. (C) TEM image of AgNCs@Co-Ni LDH nanocomposites. (D) XRD patterns comparing AgNCs@Co-Ni LDH, agarose hydrogel-patches, and hydrogel-patches. (E) UV-Vis absorption spectra of pure agarose hydrogels and hydrogel-patch. (F) FTIR profiles of nanoparticle-loaded agarose before and after MBTH functionalization. (G, H) XPS survey spectra of AgNCs@Co-Ni LDH and the hydrogel-patch, respectively. (I-L) High-resolution XPS spectra highlighting the Ag 3d, Co 2p, Ni 2p, and C 1s regions of AgNCs@Co-Ni LDH.

The optical properties of AgNCs@ZIF-67 and AgNCs@Co-Ni LDH were examined using UV-Vis absorption spectroscopy. As illustrated in **Figure S7**, AgNCs display a distinct absorption peak centered at ~400 nm, corresponding to the characteristic surface plasmon resonance (SPR) of Ag nanostructures. In AgNCs@ZIF-67, the absorption spectrum broadens across the 400-600 nm range, suggesting that the ZIF-67 shell modulates the local electronic environment of AgNCs by redistributing surface electron density, thereby altering their plasmonic behavior. Conversely, AgNCs@Co-Ni LDH exhibited a reduced absorption intensity with a less defined peak profile, likely due to the shielding effect of the LDH layers, which dampened the surface electromagnetic field of the AgNCs [32]. This reflects the synergistic interaction between the metal-organic framework and AgNCs in the core-shell structure. Additionally, the optical response of the pristine agarose hydrogel and the hydrogel-patch without MBTH was investigated. As shown in **Figure 1E**, the agarose hydrogel exhibited negligible absorbance at ~400 nm, consistent with its amorphous and optically inert nature. The hydrogel-patch without MBTH, however, demonstrated significantly enhanced absorbance, around 400 nm, primarily attributed to the SPR effect of embedded AgNCs@Co-Ni LDH. The presence of MBTH, with its conjugated aromatic system, further contributed to the observed absorption enhancement, confirming successful functionalization of the hydrogel matrix.

The FTIR spectrum of the hydrogel-patch (**Figure 1F**) displays a broad absorption band around 3300 $cm^{-1}$, attributed to O-H stretching vibrations from hydroxyl groups in the agarose matrix. Peaks at 2920 $cm^{-1}$ and 2850 $cm^{-1}$ correspond to the C-H stretching of methylene ($-CH_2-$) groups, while the band at 1620 $cm^{-1}$ arises from carbonyl (C=O) stretching vibrations. Upon incorporation of AgNCs@Co-Ni LDH, the spectrum of the hydrogel-patch retains these features while exhibiting new peaks. A distinct band at 1380 $cm^{-1}$ corresponds to $NO_3^-$ vibrations from nitrate groups in the LDH component, and the enhanced intensity near 630 $cm^{-1}$ is attributed to Ni-O and Co-O bonds vibrations, confirming the integration of metal hydroxide species [33]. Comparison with MBTH-loaded hydrogels (**Figure S8**) reveals further structural modifications. The L-hydrogel maintains characteristic agarose peaks, while the hydrogel-patch additionally shows a new band at 1380 $cm^{-1}$ due to the benzothiazole ring vibrations of MBTH, and a peak at 1580 $cm^{-1}$ is attributed to N-H bending of $-NH_2$ groups. The intensified peak around 1050 $cm^{-1}$ reaffirms the presence of AgNCs@Co-Ni LDH. These spectral changes validate the successful construction of the MBTH/AgNCs@Co-Ni LDH composite hydrogel system, confirming structural integration for targeted sensing applications.

The chemical composition and oxidation states of elements in the AgNCs@Co-Ni LDH nanocomposite and functionalized hydrogel were characterized using XPS. For the AgNCs@Co-Ni LDH composite (**Figure 1G**), spectrum confirms the presence of Ni, Co, Ag, O, and C elements, verifying the successful incorporation of all intended constituents. The characteristic Co 2p peaks, appearing at binding energies of 781.2 eV and 796.5 eV, along with the Ni 2p peaks at 855.4 eV and 873.1 eV, are accompanied by vibrational satellite peaks indicating the existence of $Co^{2+}$ and $Ni^{2+}$ species. **Figure 1H** displays the spectral profile of the hydrogel-patch, revealing prominent O 1s and C 1s peaks, which are consistent with the hydroxyl-rich, carbon-based backbone of the agarose matrix. The presence of Ag in the etched hollow composite is confirmed by the characteristic doublet peaks at 368.6 eV (Ag $3d_{5/2}$) and 374.6 eV (Ag $3d_{3/2}$), as shown in **Figure 1I**. The Co 2p spectrum (Figure 1J) exhibits main peaks at 781.2 eV and 796.5 eV, along with a corresponding satellite peak, indicating the presence of Co in the divalent state. Similarly, the Ni 2p spectrum (**Figure 1K**) shows main peaks at 855.4 eV and 873.1 eV, along with a distinct satellite peak, confirming the presence of $Ni^{2+}$ species, consistent with previous findings [33]. Furthermore, the C 1s spectrum (**Figure 1L**) supports the successful formation of the hydrogel backbone, indicating effective hydrogel network synthesis.

### 3.2. SERS performance of AgNCs@Co-Ni LDH/hydrogel-patches

The SERS activity of AgNCs@Hydrogel, AgNCs@ZIF-67@Hydrogel and AgNCs@Co-Ni LDH@Hydrogel was systematically evaluated using hexanal as a probe molecule. As shown in **Figure S9**, at a Hexanal concentration of $10^{-6}$ M, a Hydrogel containing AgNCs@Co-Ni LDH exhibits markedly higher intensity of the Raman peaks compared to AgNCs@ZIF-67 and AgNCs. This enhancement relates to the core-shell structure of AgNCs@Co-Ni LDH, where the AgNC core generates strong plasmonic resonance and the porous Co-Ni LDH shell provides abundant active sites and efficient electromagnetic field distribution, leading to dense and effective *hot-spots*. In contrast, conventional Raman spectra at the same concentration display a negligible signal, highlighting the high sensitivity of the SERS substrate. COMSOL simulations (**Figure S10**) further support these results by demonstrating enhanced electromagnetic fields surrounding the nanostructures.

The effect of etching duration on SERS performance of AgNCs@Co-Ni LDH for detecting 0.1 μM hexanal was also investigated (**Figure S11**). Signal intensity gradually increases with etching time, reaching a maximum at 5.5 h, where characteristic hexanal peaks at ~850 $cm^{-1}$ and 1250-1300 $cm^{-1}$ are most distinct. Shorter etching times yield weaker signals, while over-etching (6 h) leads to signal deterioration due to structural degradation. Thus, 5.5 h was identified as the optimal etching time for preparing AgNCs@Co-Ni LDH-functionalized hydrogel-patches for gas sensing applications. To further validate the experimental observations, DFT calculations were conducted to simulate the Raman spectra of the Hexanal-MBTH derivative using the B3LYP functional and the 6-31G(d,p) basis set. The theoretical spectra (**Figure S12A- S12C**) exhibit strong agreement with the experimental SERS data, thereby confirming the vibrational assignments and elucidating the molecular interactions responsible for the observed spectral features. Furthermore, geometry optimizations of MBTH, FA-MBTH, and Hexanal-MBTH derivatives were performed (**Figure S12D- S12F**), providing detailed insights into their structural configurations and binding mechanisms. Following the theoretical analysis, condition optimization experiments were carried out on the hydrogel-patches. As shown in **Figure S13A-S13B**, the SERS performance is optimal under ambient conditions with an MBTH concentration of 1.0 mg/mL and a reaction time of 20 min. Furthermore, **Figure S13C-S13F** demonstrate that neither the degree of agarose crosslinking nor the pH of the hydrogel had a significant effect on the SERS signal intensity.

### 3.3 Qualitative dual-mode detection via SERS and colorimetric approaches

The influence of hexanal on the SERS spectra of the hydrogel-patch is depicted in **Figure 2A**. Upon exposure to hexanal, we see a significant enhancement in the SERS peak at 1277 $cm^{-1}$ (**Figure 2B**), which corresponds to the formation of hexanal-MBTH adducts via $Fe^{3+}$-mediated oxidative coupling [34]. This characteristic peak arises from the superimposed vibrational contributions of ν(C–N) (C–N stretching), ρ(CH) (CH out-of-plane bending), δ(N–C–N) (N–C–N in-plane bending), and ν(N–C–N) (N–C–N stretching) modes within the hexanal-MBTH complex. In the absence of hexanal, only a weak signal at 1277 $cm^{-1}$ was detected, primarily attributable to the intrinsic ν(N–C–N) vibrations of MBTH. Furthermore, comparative Raman spectra of pure MBTH and its aqueous mixture with hexanal exhibits clear differences in the 1200-1400 $cm^{-1}$ region, supporting the generation of an azine-type derivative characterized by the formation of an "N-N-CH-N-N" bridging motif. The Raman spectra of the reaction intermediates and the final hexanal-MBTH derivatives exhibit significant differences, which are likely attributable to distinct adsorption behaviors arising from structural variations between the hydrogel-supported state and the dried state. Importantly, the SERS signals predominantly originate from derivative molecules adsorbed onto the active substrate via their aromatic ring moieties. The measured polarizabilities of MBTH and MBTH-hexanal were 116.7 and 487.4, respectively, indicating that higher molecular polarizability significantly enhances Raman signal intensity.

To exclude the influence of residual substances in the synthesis steps on the results, samples from each step (AgNCs, AgNCs@ZIF-67, agarose hydrogel, AgNCs@Co-Ni LDH, hydrogel-patch) were taken in certain quantities and hexanal was added. Significant colorimetric changes are observed exclusively in the final AgNCs@LDH samples (**Figure 2C**), indicating that preceding synthesis steps does not interfere with the dual-mode detection performance. The chemical reaction scheme in **Figure 2C** underscores the specificity of the detection method and highlights its applicability for trace gas adsorption. The hydrogel-patch exhibits a distinct colorimetric response to increasing concentrations of aldehyde gas (from 0 M to $10^{-6}$ M), transitioning gradually from yellow to blue. This color shift enabled effective quantitative detection using a smartphone-based application (Color Grab). Time-dependent imaging (**Figure 2D**) of the hydrogel-patch further confirmed that the intensity of the color change correlated positively with hexanal exposure duration, supporting its potential for real-time aldehyde monitoring.

The integration of sensing components into a hydrogel matrix offers several critical advantages: (1) facile replaceability and compatibility with wearable platforms, such as face masks, thereby enhancing practical deployment; (2) enhanced stabilization of AgNCs@Co-Ni LDH nanoparticles by the agarose matrix, which effectively minimizes nanoparticle aggregation; (3) increased adsorption efficiency due to the high specific surface area of the hydrogel, facilitating effective hexanal uptake; (4) establishment of a conducive microenvironment for rapid derivatization reactions, enabling the porous agarose network and optimizing MBTH concentration; and (5) superior environmental adaptability, characterized by excellent portability, storage stability, and performance robustness across varied gas sampling conditions.

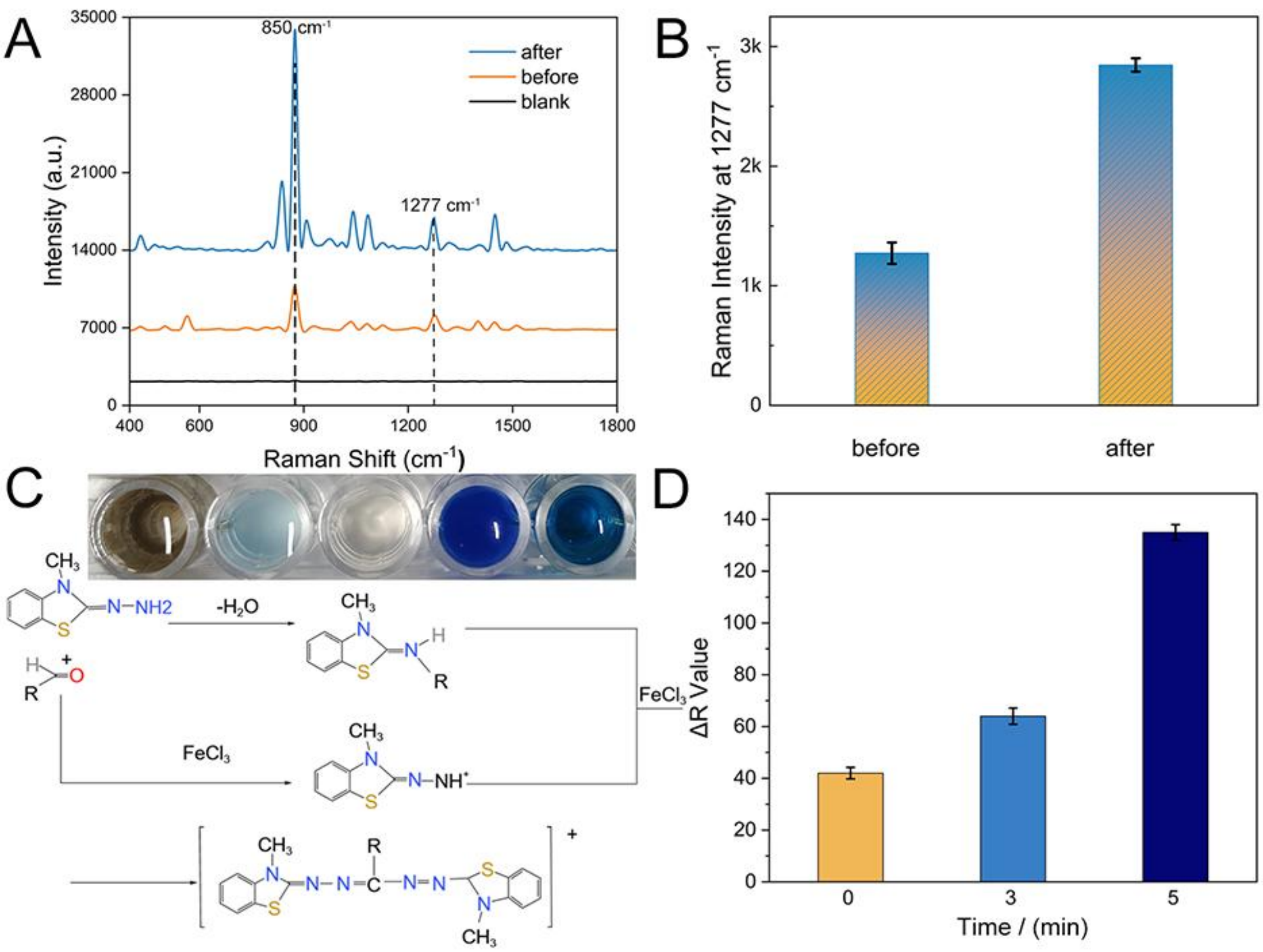

**Figure 2.** (A) Raman spectra recorded on hydrogel-patch before and after the addition of hexanal, along with the control group without MBTH. (B) Intensity change map of the hexanal characteristic peak at 1277 $cm^{-1}$ before and after the reaction. (C) Color photograph of the reaction of hexanal with the synthesized material at each step. (D) Schematic diagram showing color changes on hydrogel-patch patches before and after capturing hexanal (0.5 $mg/m^3$) at 0 min, 3 min, and 5 min.

**Figure S14** highlights the minimal variation in RGB values recorded across multiple smartphone models, demonstrating the high inter-device compatibility and reliability of the proposed colorimetric sensing platform. Such consistency underscores the system's robustness and suitability for deployment in diverse, real-world diagnostic settings. **Figure S15** presents a custom-engineered, 3D-printed smartphone attachment specifically designed to standardize image acquisition in portable sensing applications. This attachment ensures precise alignment of the hydrogel-patch and effectively mitigates ambient light interference, thereby facilitating accurate and reproducible RGB data capture across various smartphone devices.

### 3.4 Quantitative detection of Hexanal using SERS and colorimetric dual-mode

To improve the detection sensitivity of hexanal, an air bag-assisted sampling approach was employed in conjunction with a smartphone-based colorimetric analysis platform. **Figure 3A** illustrates the optical image analysis for hexanal detection and the methodology for establishing the concentration-color card. Following image acquisition via a smartphone, an application "Color Grab" was used to extract RGB values from each sample by selecting a fixed circular region (diameter: 6.4 mm), thereby ensuring consistency in signal analysis. The resulting RGB data enabled the generation of a reliable colorimetric calibration curve, exhibiting a distinct color gradient with a limit of detection (LOD) as low as $1\times10^{-6}$ M. To further assess the colorimetric response of the hydrogel-patch, optical images of the reaction solutions captured by the smartphone's built-in camera were analyzed to quantify hexanal concentrations. An intensification of the color signal is observed, as hexanal concentrations increased from $1\times10^{-12}$ M to $1\times10^{-7}$ M. It is attributable to the specific reaction between MBTH and hexanal (**Figure 3B, inset**). A strong linear relationship is observed between the color intensity and the logarithm of hexanal concentration ($R^2 = 0.99$), with a calculated LOD of $1.26\times10^{-11}$ M (LOD = 3σ/slope, where σ represents the standard deviation of the blank, S/N = 3). Compared to existing hexanal detection technologies, our platform demonstrates superior sensitivity and a broad detection range (**Table S1**). To facilitate semi-quantitative visual detection, a concentration-color reference chart was constructed, correlating color intensity with hexanal concentration and enabling reliable naked-eye evaluation. The accuracy of this method was validated by comparing the results with Raman spectroscopy, confirming the reliability of our dual-mode detection approach (**Figure 3C**). Raman spectroscopy of the corresponding samples reveal a pronounced increase in signal intensity with rising hexanal concentrations, ranging from $1\times10^{-12}$ M to $1\times10^{-8}$ M (**Figure 3D, inset**). The close agreement between SERS and colorimetric results underscores the accuracy, robustness, and versatility of the dual-mode sensing platform for sensitive hexanal detection.

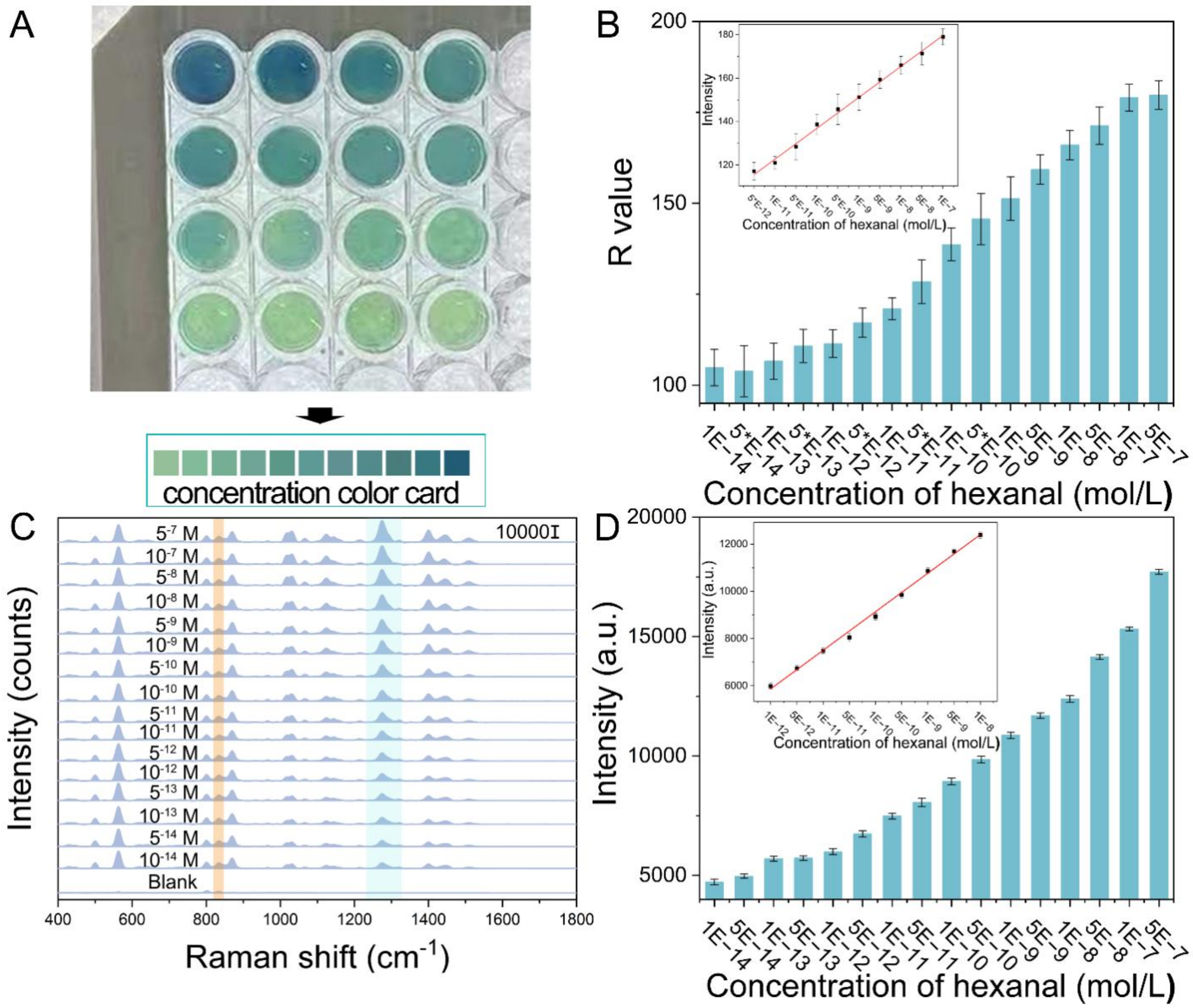

**Figure 3**. (A) Real image showing the color change after adding different concentrations of hexanal. (B) Quantitative histogram of colorimetric detection for hexanal. (C) SERS signal spectrum showing changes at different hexanal concentrations. (D) SERS signal change at 1277 $cm^{-1}$ for varying concentrations of hexanal.

### 3.5 Resistance analysis of the hydragel patch

The signal uniformity of the hydrogel-patch was determined by conducting SERS analysis at 11 randomly chosen points across the patch surface following hexanal exposure. The intensity of the characteristic peak at 1277 $cm^{-1}$ was quantified, resulting in a relative standard deviation (RSD) of 2.7%, which demonstrates excellent signal homogeneity and repeatability of detection across the patch surface (**Figure 4A**). Moreover, the anti-interference performance of the hydrogel-patch was explored by examining its response to potential interfering compounds using SERS analysis. Common volatile organic compounds present in the exhaled breath of lung cancer patients including acetaldehyde, benzoic acid, propionaldehyde, and ethanol were selected as representative interferents. The SERS spectra shown in **Figures 4B-4C** reveal a well-defined hexanal MBTH labeling peak at 1277 $cm^{-1}$, demonstrating the superior selectivity of the hydrogel-patch for hexanal detection over these interfering compounds. Another key diagnostic feature for hexanal-MBTH compound is the prominent ~850 $cm^{-1}$ peak, assigned to C-C skeletal stretching vibrations of the 6-carbon straight chain in the hexanal-MBTH adduct. This band is absent or negligible in short-chain aldehydes, which instead exhibits C-C skeletal vibrations below 800 $cm^{-1}$ (e.g., acetaldehyde at ~780 $cm^{-1}$).

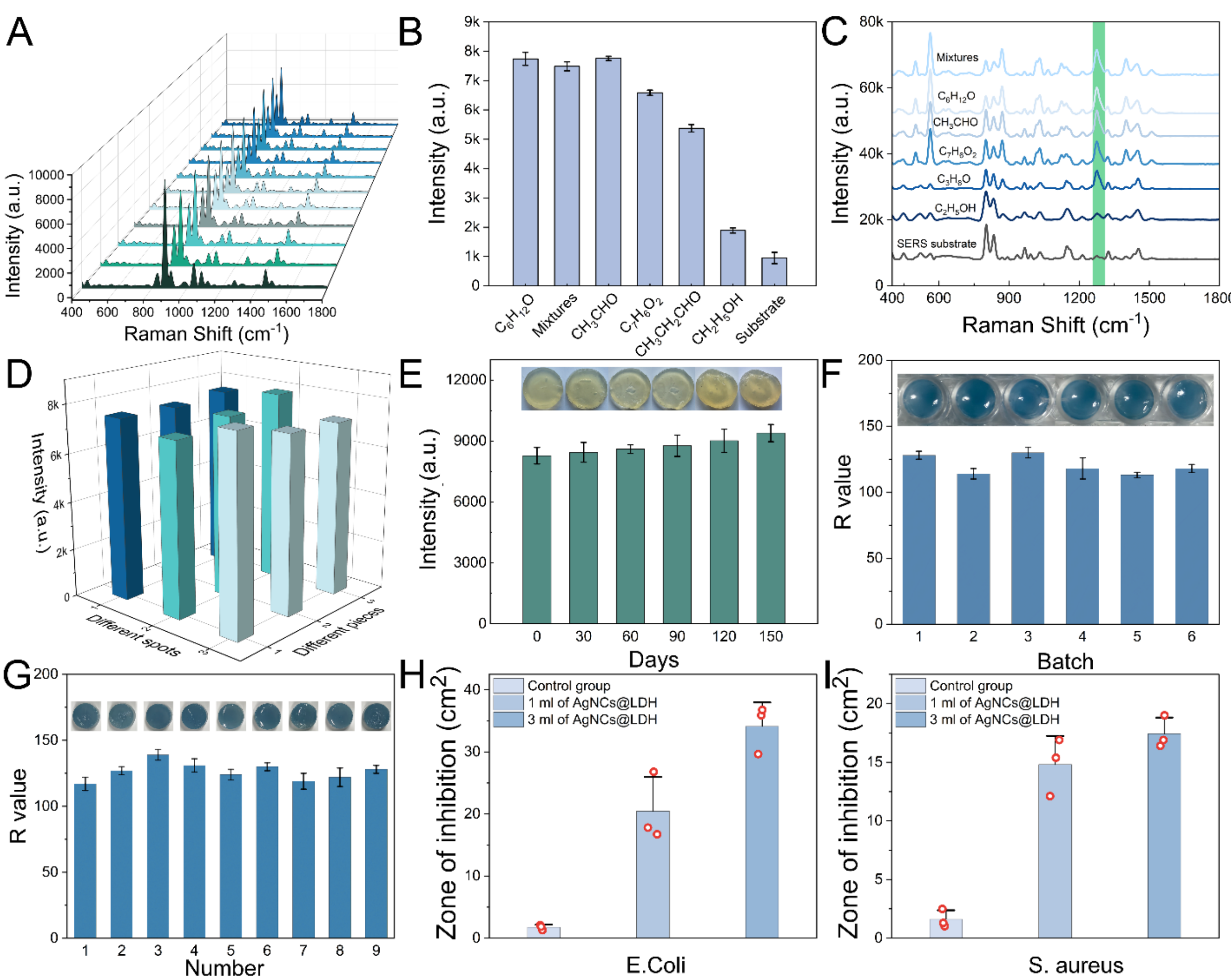

**Figure 4**. (A) SERS spectra from 11 random points on an hydrogel-patch exposed to 0.05 mg/$m^3$ hexanal. (B) Histogram of the hydrogehydrogel-patch in the presence of hexanal and other coexisting substances. (C) Raman analysis of the hydrogel-patch's resistance to against various potential interfering components. (D) SERS

intensity of 0.05 $mg/m^3$ hexanal gas were obtained from three batches of patches and three different acquisition points on each patch. (E) Raman peak intensity at 1277 $cm^{-1}$ after 0 to 150 days of storage (inset: corresponding physical photos). (F) Signal intensity of R-values for different batches of patches. (G) R-value signal intensity for different patches in the same batch. (H) Statistical graph of the sterilization zone area of the hydrogel-patch against *Escherichia coli*. (I) Statistical graph of the sterilization zone area of the hydrogel-patch against *Staphylococcus aureus*.

To further investigate the reproducibility of the preparation process, SERS intensity values at 1277 $cm^{-1}$ were measured across three different batches of patches, with three sampling points selected per batch. Data in **Figure 4D** reveal excellent manufacturing reproducibility, with an RSD of 3.86 % at 1277 $cm^{-1}$, indicating consistent quality across multiple production batches. The temporal stability of the hydrogel-patch matrix was monitored over a 150-days period under ambient storage conditions. Raman spectra acquired at predefined intervals (**Figure 4E**) reveal a negligible signal degradation, with a minor increase in SERS intensity over time, plausibly due to slow water loss via evaporation. This slight enhancement in signal intensity, despite long-term storage in sealed Petri dishes, underscores the physicochemical stability of the hydrogel framework. Importantly, this stability mitigates the risk of false-negative or false-positive outcomes during extended deployment and enhances the practicality of the platform for real-world applications. Time-stability measurements for colorimetric detection (**Figure S16**) showed that the R-value increased initially, reaching a maximum at 6-12 h, and then gradually declined due to oxidation. The signal remained for at least 48 h, demonstrating the colorimetric method's suitability for offline or delayed hexanal analysis. Additionally, the stability and reproducibility of the colorimetric response were evaluated at both the inter-batch and intra-batch levels to further substantiate the analytical reliability of the platform. R-value signal intensities derived from different batches are shown in **Figure 4F**. Despite minor fluctuations, the mean R-values across batches remained statistically identical, suggesting minimal batch-to-batch variability and confirming the reproducibility of the fabrication process. This consistency reflects tight control over material formulation and chromophore distribution during synthesis. Furthermore, the R-value signal intensity of individual patches within the same batch exhibited low dispersion and remained relatively stable, further confirming the high reliability and consistency of the patch quality within each batch, **Figure 4G**.

Finally, considering the intended application of the hydrogel-patch for respiratory gas detection, its antibacterial performance was systematically evaluated. The antimicrobial activity of AgNCs@Co-Ni LDH incorporated into agarose hydrogel-patches was assessed against two representative bacterial strains: *Escherichia coli* (Gram-negative) and *Staphylococcus aureus* (Gram-positive), across varying concentrations. As shown in the real images (**Figure S17-S18**), robust bacterial proliferation was observed on the agar plate in the control group, whereas zones surrounding the AgNCs@Co-Ni LDH-containing patches exhibited minimal to no bacterial growth. Quantitative analysis of the inhibition zones (**Figures 4H-4I**) demonstrate that both 1 mL and 3 mL hydrogel-patches effectively suppress bacterial growth, with the larger-volume (3 mL) patches producing significantly wider zones of inhibition. These results confirm dose-dependent antibacterial activity of the AgNCs@Co-Ni LDH composite against both bacterial species [35]. The observed efficacy is primarily attributed to the intrinsic antimicrobial properties of AgNCs, which are known to disrupt bacterial membrane integrity, interfere with essential intracellular processes, and inhibit cellular proliferation. One of the key antibacterial mechanisms of AgNCs involves the induction of potassium ion efflux from bacterial cells, leading to cellular dysfunction and death [36]. In the composite system, Co-Ni LDH matrix likely functions as a stable and efficient support for AgNCs immobilization and controlled release, thereby enhancing their bioavailability and interaction with bacterial cells. These findings highlight potential of AgNCs@Co-Ni LDH agarose hydrogel-patches in antimicrobial applications, particularly in contexts requiring integrated biosensing and infection control.

## 3.6 Analysis of exhaled gas samples from clinical lung cancer patients

To evaluate the practical applicability of the hydrogel-patch for clinical detection, we developed three distinct methodologies based on MBTH-AgNCs@Co-Ni LDH for rapid detection using this hydrogel-patches. The first method involves connecting an air pump to a Tedlar air bag for the collection of respiratory gases from suspected lung cancer patients (**Figure S19**). In this configuration, the sampling bag was linked to the sensor inlet, while a micro air pump is connected to the outlet. The pumping rate was maintained at 25 mL/min, and after continuous pumping for 5 min, the collected samples are directly subjected to both SERS and colorimetric detection. The second approach integrates the hydrogel onto a 300-mesh stainless steel mesh (**Figures 5A-5B)**. The mesh-hydrogel assembly was subsequently connected to a syringe and gas actively delivered through the patch by pressing the plunger. The third method incorporates the hydrogel-patch into a mask, allowing detection to occur through natural breathing by the patient while wearing the mask (**Figure S20**). The mask-based sensing platform offers multiple advantages, particularly in terms of user comfort, portability, and cost-efficiency. As shown in **Table S2**, the total mass of the breath-detecting mask is only 8.94 g, including the 3D-printed valve and sensing components, ensure lightweight wearability suitable for prolonged clinical use. Moreover, **Table S3** reveals that the total material cost per hydrogel-patch sensor is approximately 0.41$, demonstrating the economic feasibility of large-scale production.

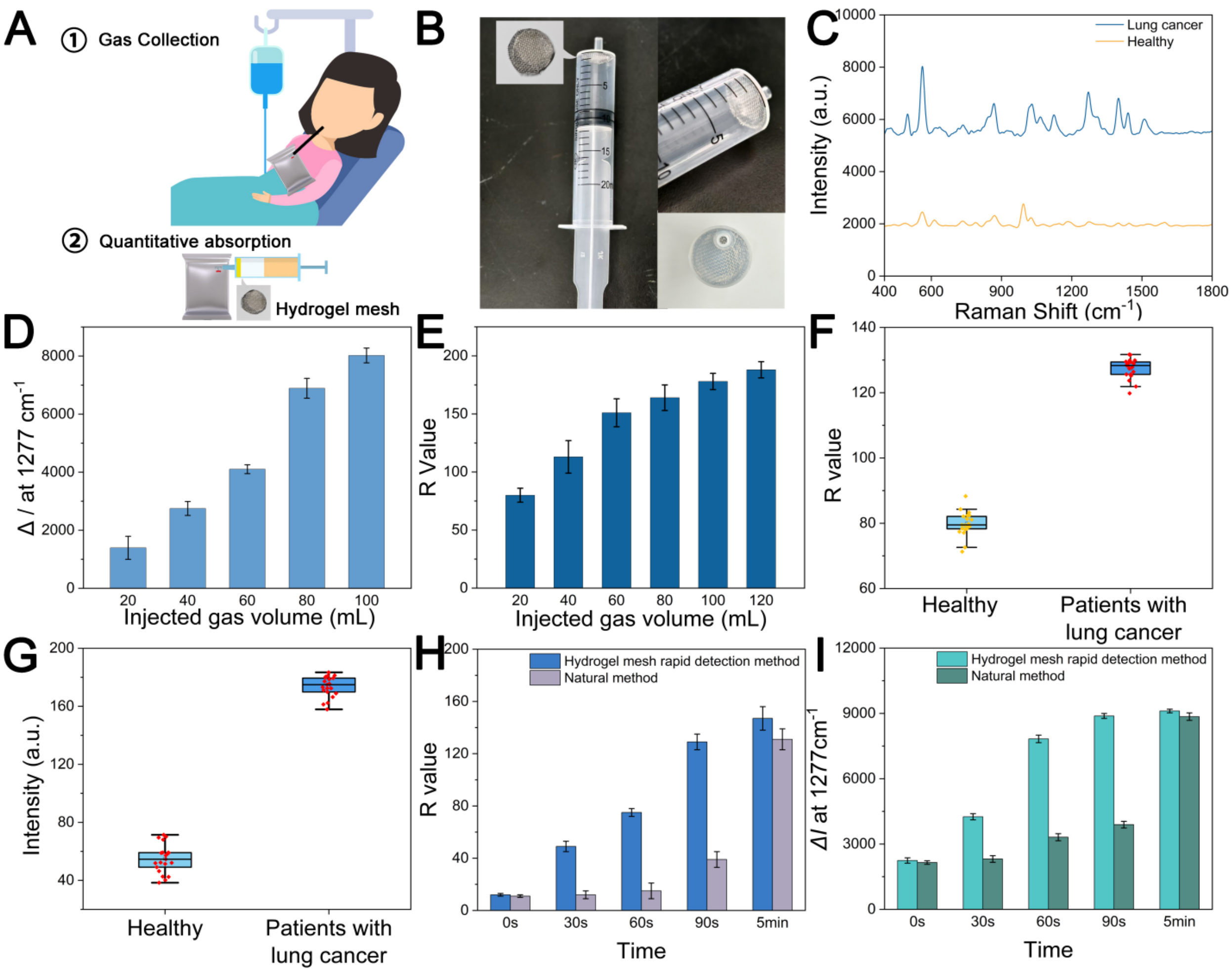

**Figure 5**. (A) Schematic diagram of the clinical gas collection and detection. (B) Real image of exhaled gas from lung cancer patients using the hydrogel on stainless steel mesh. (C) Average SERS spectra of healthy and lung cancer patients. (D) SERS signal at 1277 $cm^{-1}$ across different breath volumes. (E) R-value changes with varying hexanal concentrations. (F-G) Boxplots of SERS signal and R-value at 1277 $cm^{-1}$ for exhaled gas from 20 lung cancer patients and 20 healthy individuals. (H) R-values at different time points using hydrogels loaded with stainless steel mesh for rapid detection and the natural method. (I) Schematic diagram of 1277 $cm^{-1}$ SERS signal at different time points for both detection methods.

Upon analysis of actual samples, the second detection method, which utilizes the hydrogel mesh-syringe system, demonstrated the highest efficiency, reducing the detection time to approximately 1.5 min. Furthermore, this hydrogel mesh-syringe method is cost-effective and features a simple, user-friendly operation, making it an attractive option for practical applications in clinical settings. A comparative evaluation of detection time requirements across various application modes and analytical methodologies is summarized in **Table S4**, further emphasizing the superior responsiveness and practical applicability of the hydrogel mesh-syringe system in diverse detection scenarios. Meanwhile, the practical potential of the MBTH-based AgNCs@Co-Ni LDH detection approach was demonstrated by evaluating the ability of hydrogel-patches to distinguish respiratory gases from lung cancer patients and healthy individuals. The average SERS spectra for healthy and lung cancer patients were obtained by analyzing the collected exhaled gases (**Figure 5C**), which revealed marked differences between the two groups. **Figure 5D** illustrates the variation in SERS signal intensity at 1277 $cm^{-1}$ as a function of the volume of exhaled gas, indicating a positive correlation between signal intensity and gas volume. **Figure 5E** further demonstrates the variation in R-value with exposure to different concentrations of hexanal, highlighting the method's sensitivity to changes in hexanal concentration. A comparative analysis of exhaled gases from 20 lung cancer patients and 20 healthy individuals reveals distinct differences in both SERS signal intensity and R-value signal intensity at 1277 $cm^{-1}$, as depicted in the boxplots in **Figures 5F-5G**. **Table S5** summarizes the clinical characteristics of 20 lung cancer patients, including age, gender, smoking status, and cancer pathology. The data show that the SERS signal intensity and R-value at 1277 $cm^{-1}$ are significantly higher in lung cancer patients than in healthy individuals, thereby facilitating the differentiation between the two groups. Furthermore, the effects of rapid detection using hydrogel-loaded stainless-steel mesh, compared with the natural breathing method at various time points, were also evaluated. **Figure 5H** presents the bar graph of R-values at different time intervals, while **Figure 5I** provides a schematic representation of the SERS signal at 1277 $cm^{-1}$. These results suggest that the hydrogel-based rapid detection method offers distinct advantages in terms of detection efficiency and signal response, providing valuable insights for the development of rapid clinical detection for lung cancer. Moreover, in comparison with other reported hexanal biosensors, the dual-mode sensor analysis demonstrates significant advantages, including superior sensitivity, high accuracy, and ease of use, positioning this method as a promising tool for clinical applications.

### 3.7 Deep learning-based portable image recognition system

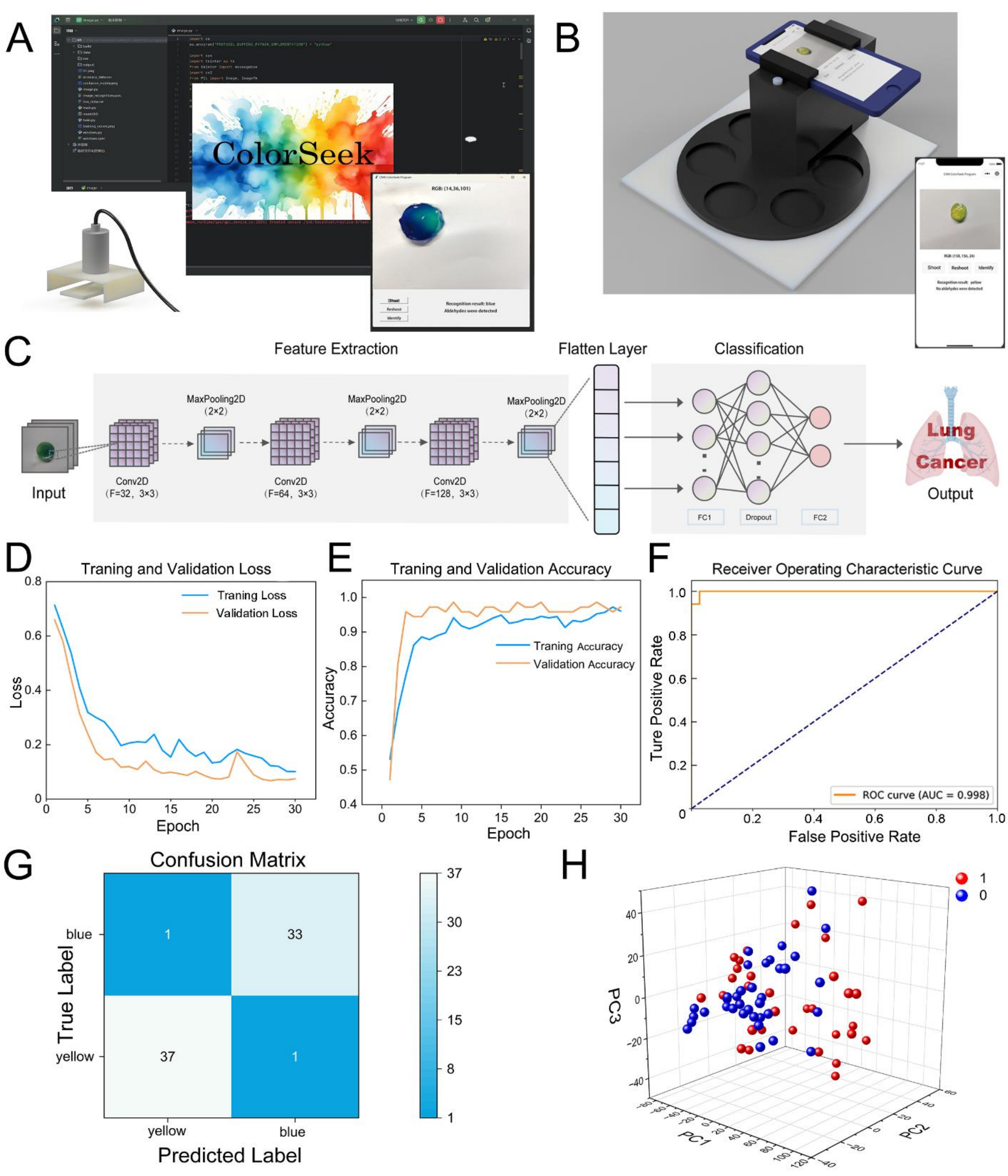

**Figure 6.** Workflow and performance of the ColorSeek convolutional neural network (CNN) system. The application operates on PC (**A**) and smartphone (**B**) platforms. The CNN architecture (**C**) comprises convolutional and max-pooling layers for feature extraction, followed by flattening and fully connected layers for lung cancer classification. After over 30 training epochs, the model's loss decreased (**D**) and accuracy exceeded 95% (**E**). The receiver operating characteristic (ROC) curve achieved an area under the curve (AUC) of 0.9998 (**F**). The confusion matrix (**G**) summarizes classification performance, while the principal component analysis (PCA) plot (**H**) shows limited class separation, underscoring the accuracy and robustness of the CNN-based approach.

After confirming the feasibility of the dual-mode detection hydrogel based on AgNCs@Co-Ni LDH for hexanal detection, we further developed a multi-terminal image recognition program using deep learning to enable rapid on-site analysis of hexanal in human breath samples. The described multi-terminal recognition program is illustrated in **Figures 6A** and **6B**, which depict the recognition processes using PC-based and smartphone-based applications, respectively. On both platforms, the model successfully identified the hydrogel types and corresponding RGB values, and this advancement supports the rapid and non-invasive detection of lung cancer. The complete image recognition workflow is shown in **Figure 6C** and follows the algorithmic framework detailed in this **Section**. hydrogel-patches were exposed to exhaled breath samples from two cohorts: lung cancer patients and healthy individuals. Each patch was photographed under standardized lighting conditions using a fixed photography mold to ensure uniformity. Two images were captured per sample, and the one with better clarity was selected. These curated images were used to construct a dataset for deep learning-based feature extraction and classification. The training and validation results of the CNN model are presented in **Figures 6D** and **6E**, with an accuracy exceeding 95% and a loss value below 0.1. Meanwhile, the AUC value of the ROC curve of this model is as high as 0.998, as shown in **Figure 6F**, which is almost close to the upper left corner, indicating that the model has extremely excellent classification performance and significant ability to distinguish between positive and negative samples. The confusion matrix in **Figure 6G** further confirms the high classification performance of the model. For comparison, principal component analysis (PCA) was performed, as shown in Figure 6H. However, PCA did not yield clear class separation, thereby highlighting the superior accuracy and robustness of the CNN-based approach. To evaluate practical applicability, the trained model was tested on hydrogel-patches outside the original dataset. These findings demonstrate that the developed multi-terminal recognition system can accurately and portably detect lung cancer-related biomarkers, offering a promising tool for point-of-care diagnostics.

## 4. Conclusion

In conclusion, a convenient SERS-based hydrogel-patch was developed for the on-site selective detection of hexanal, leveraging the specific reaction between hexanal and MBTH, along with the high adsorption capacity of the hydrogel matrix. SERS analysis demonstrated an ultralow LOD of $3.34\times10^{-13}$ mg/mL, while colorimetric analysis achieved a wide nearly linear detection range of from $1\times10^{-12}$ M up to $1\times10^{-7}$ M. Furthermore, CNN based portable image recognition system was also developed to automatically analyze and quantify the colorimetric results to facilitate real-time, user-friendly interpretation particularly in non-laboratory or field settings. The overall detection platform, comprising a functionalized hydrogel-patch mounted on a stainless-steel mesh and operable via a syringe system, demonstrated strong potential for rapid, on-site monitoring of trace hexanal in exhaled breath. This integrated strategy offers high specificity, sensitivity, operational simplicity, and reliability, making it a promising tool for early screening and assessment of suspected lung cancer patients.

## Acknowledgments

J.L, S.W and A.A. contributed equally to the work. The research was supported by the National Natural Science Foundation of China (Grant No. 11635013 and T2250410383) and Anhui Provincial Key Research and Development Plans (202004i07020014). Additional support was provided Chinese Academy of Sciences Special Exchange Program 2025.